\documentclass{an}

\usepackage{graphicx}
\usepackage{times}
\usepackage{fancyhdr}
\begin{document}

\title{Drop of coherence of the lower kilo-Hz QPO in neutron stars: is there a link with the innermost stable circular orbit?}

\author{Didier Barret$^{1}$\thanks{E-mail: Didier.Barret@cesr.fr},
Jean-Francois Olive$^{1}$ and M. Coleman Miller$^{2}$}
\institute{$^{1}$Centre d'Etude Spatiale des Rayonnements, CNRS/UPS,
9 Avenue du Colonel Roche, 31028 Toulouse Cedex 04, France \\
$^{2}$Department of Astronomy, University of Maryland, College Park,
MD 20742-2421, United States}
\date{Received; accepted; published online}

\abstract{Using all available archival data from the Rossi X-ray
Timing Explorer (RXTE), we follow the frequency of the kilo-Hz QPOs
in three low luminosity neutron star low mass X-ray binaries; namely
4U~1636-536, 4U~1608-522, and 4U~1735-44. Following earlier work by
Barret et al. (2005a,b), we focus our analysis on the lower kilo-Hz
QPO, for which we study the dependency of its quality factor
(Q=$\nu/\Delta\nu$, where $\Delta\nu$ is the FWHM) and amplitude as a function
of frequency over a range covering from 500 Hz to 1000 Hz. As
previously found for 4U~1636-536, we show that the quality factor of
the lower kilo-Hz increases with frequency up to a maximum frequency
around 800 Hz, beyond which an abrupt drop of its coherence is
observed down to a limiting frequency where the QPO disappears
completely. Simultaneously the amplitude of the QPOs is almost
constant below the peak frequency and starts to decrease smoothly
afterwards. The peak frequency is 850 Hz, 820 Hz, 740 Hz whereas the
limiting frequency is 920 Hz, 900 Hz and 830 Hz  for 4U~1636-536,
4U~1608-522 and 4U~1735-44 respectively. A  ceiling of the lower QPO
frequencies is also seen clearly in a frequency versus count rate
diagram for all sources. This behavior is reproducible within an
object and between objects.  We suggest here that the drop of
coherence of the lower QPO may be a geometry-related effect, which could
be related to the last stable circular orbit.
\keywords{Accretion - Accretion disk, stars: individual 4U~1636-536,
4U~1608-522, 4U~1735-44, stars: neutron,  stars: X-rays}}

\correspondence{barret@cesr.fr}
\maketitle
\section{Introduction}
Using archival data from the Rossi X-ray Timing Explorer (RXTE), we have
studied in a systematic way the variation of the quality factor
(Q=$\nu/\Delta\nu$, where $\Delta\nu$ is the FWHM) and amplitude of the
lower and upper kilo-Hz quasi-periodic oscillations (QPO) in the low-mass
X-ray binary 4U~1636-536.  It has been shown that the lower and upper QPO
follow two different tracks in a frequency versus quality
factor, as represented. The lower QPO is a relatively broad feature with
Q $\sim$ 20 at $\sim$ 600 Hz. Its quality factor then increases steadily
up to $Q\sim 200$ at $\sim$850 Hz, beyond which it drops precipitously to
$Q\sim 30$ at the highest detected frequencies $\sim 920$~Hz. The drop of
coherence is accompanied by a smooth drop of amplitude. A ceiling of the
QPO frequency at the latter frequency is also clearly seen in a
frequency-count rate diagram. The behavior of the lower QPO of 4U~1636-536 is summarized
in Figure 1.

On the other hand, the upper QPO shows a clear positive correlation between
its quality factor and its frequency, with no evidence for a drop when it
reaches its highest frequencies. The amplitude of the upper QPO decreases
steadily with frequency, bringing the signal down to the sensitivity level of
our analysis, and therefore making the measurement of its quality factor
difficult and certainly subject to biases (only the narrower signals are
seen, see Barret et al. 2005b for details).

Saturation of frequency and drops of amplitude and coherence of kilo-Hz QPOs
were anticipated as possible signatures of the innermost stable circular
orbit (ISCO) around a sufficiently compact neutron star (Miller et al.
1998). It was however thought that this should be seen firstly for the
upper QPO, which in most models, is associated with orbital motion. 

In this paper, we present the results of an analysis of the lower
QPO  in two additional sources: 4U~1608-522 and 4U~1735-44, using all the publicly available RXTE archival
data as of July 2005. We have
selected 4U~1608-522 because in a rather limited data set, a loss of
coherence of the lower QPO was reported by Berger et al. (1996) and
Barret et al. (2005a). The choice of 4U~1735-44 is motivated by the
fact that not much is known about its kilo-Hz QPOs (see however
Wijnands et al. 1998 and Ford et al. 1998), while the source has been rather extensively observed with RXTE.

\section{Data Analysis}
The analysis is performed the exact same way as in Barret et al. (2005b). We
have retrieved all the Proportional Counter Array science event files
with time resolution better than or equal to 250 micro-seconds. Only
data files with exposure times larger than 600 seconds are considered
here. For 4U~1608-522, the data set covers the period from March 1996
to March 2004, whereas for 4U~1735-44 the data cover from August 1997
to September 1999. No filtering on the raw data is performed, which
means that all photons are used in the analysis, and only time intervals
containing X-ray bursts are removed. We have computed Leahy normalised
Fourier power density spectra (PDS) between 1 and 2048 Hz over 8 s
intervals (with a 1 Hz resolution).

Because at their lowest end, QPOs are broad features (with Q of a few),
the analysis is restricted to QPOs detected above 500 Hz. The first stage
analysis intends to estimate the Q the strongest QPO by
removing as much as possible the long term frequency drift within the
segment. This is achieved through a shift-and-add technique performed on
the shortest possible timescales permitted by the data statistics. In this
analysis, the possible timescales considered are 16, 32, 64, 128, 256,
512 and 1024 seconds. The first product of our analysis is therefore the
mean parameters of the strongest QPO, which is fitted with a Lorentzian,
above a fitted counting noise level. In some data files, two QPOs can be
detected, allowing direct identification of the strongest one. In
4U~1608-522, it is generally the lower QPO which is the strongest of the
two. This is no the case for 4U~1735-44. As previously found in Barret et
al. (2005b), in those segments where only one QPO is detected, the
position of the QPO on a quality factor versus frequency plot allows its
identification. For 4U~1636-536, we have shown that QPOs with Q larger than
$\sim 20$ in the range 650 Hz to 950 Hz are lower QPOs. For 4U~1608-522, lower QPOs have a minimum Q of $\sim 50$ between 600 and 900 Hz. For 4U~1735-44 a limiting Q of $\sim 30$ is inferred. In Figures 2 and 3,
we show the dependency of the quality factor and amplitude of the
identified lower QPOs of 4U~1608-522 and 4U~1735-44. These two figures show
that the behavior seen from 4U~1636-536 is also present in these two
objects; mostly a drop of coherence at a critical frequency. The properties of the upper QPOs will be described in a forthcoming
paper, but as in 4U~1636-536, we have not found clear evidence of a drop of
coherence at the highest detected frequencies. Note that the maximum Q value observed for 4U~1608-522 is remarkably similar to the one of 4U~1636-536 (it is significantly smaller for 4U~1735-44).

\begin{figure}
\resizebox{\hsize}{!}
{\includegraphics[]{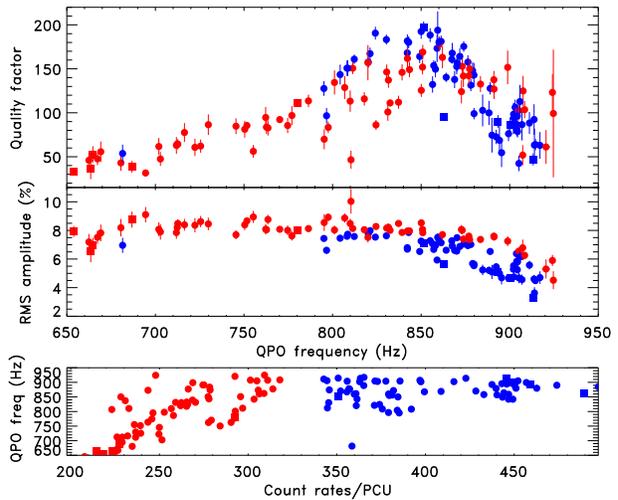}}
\caption{The quality factor of the lower kilo-Hz QPO in 4U~1636-536
showing a smooth rise with frequency up to a critical frequency
beyond which a rather abrupt drop is observed. The colors indicate
the separation between the observations performed at count rate
below (red) and above (blue) the mean count rate of all the
observations. The filled squares are those points for which an
upper QPO was detected above the $2\sigma$ level in the same
segment. The mid panel shows the dependency of the amplitude of
the lower QPO with frequency. The amplitude remains more or less
constant up to the critical frequency and then decreases smoothly
afterwards. The bottom panel shows the frequency against the mean
count rate within a continuous segment of data. Whatever the count
rate is, the QPO frequency does not exceed a limiting frequency.  This
figure shows that the frequency, rather than the countrate, is the
governing parameter for the properties of the lower QPO.}
\label{figlabel}
\end{figure}
\begin{figure}
\resizebox{\hsize}{!}
{\includegraphics[]{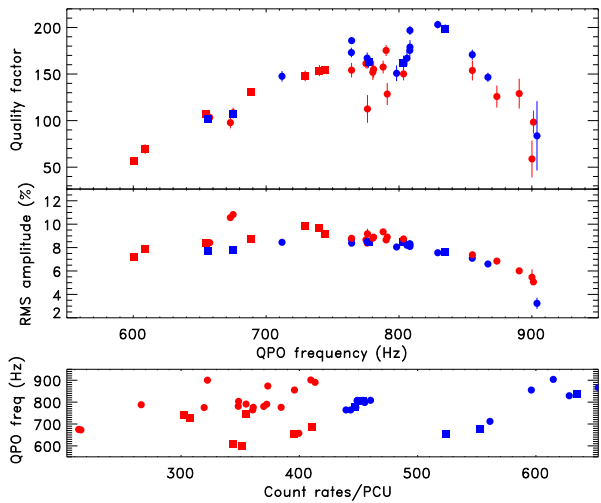}}
\caption{Same as figure 1 for 4U~1608-522.}
\label{figlabel}
\end{figure}
\begin{figure}
\resizebox{\hsize}{!}
{\includegraphics[]{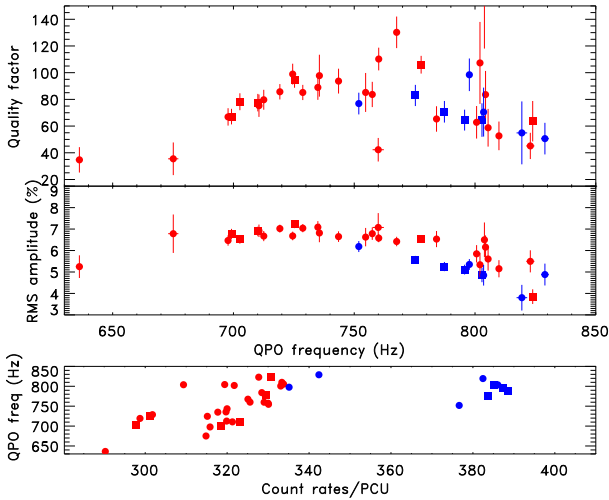}}
\caption{Same as figure 1 for 4U~1735-44.}
\label{figlabel}
\end{figure}

\section*{Discussion}

The rapid drop seen in the quality factor of the lower QPO in 4U~1636--536,
4U~1608--52, and 4U~1735--44 occurs reproducibly at a particular
frequency for each source: the peak frequency is 850 Hz, 820 Hz, 
and 740 Hz for 4U~1636-536, 4U~1608-522 and 4U~1735-44 respectively.   At the
same time, the ``parallel tracks" phenomenon (e.g. M\'endez et al. 2001)
means that a given frequency can be reached at many different
countrates.  We therefore observe a phenomenon that appears to depend
primarily on frequency rather than countrate.  Indeed, as Figures 1 to
3 show, the dependence of quality factor on frequency, and of rms
amplitude on frequency, is consistent between the high and low
countrate samples.  In addition, the saturation frequency is the same
over a factor of two in countrate.  What does this imply about the
mechanism that decreases the quality factor?

If the clock that determines the QPO frequencies originates in the
disk and is related to orbital, epicyclic, or beat frequencies, then
the sharp drop in $Q$ at a fixed frequency implies that there is some
particular radius in the disk inside of which the mechanism that
generates QPOs loses coherence.  There are several different factors
that contribute to the sharpness of the QPO. For example the quality factor depends on
(1)~the range of radii over which the QPOs are generated,
(2)~the number of cycles that the QPO lasts, and (3)~the
amount by which the radius of QPO generation drifts radially during its lifetime.  
In addition, assuming that most of the energy in the QPO is actually 
released upon impact with the stellar surface, changes in the 
dynamics of the boundary layer could affect the quality factor.

With this in mind, it could in principle could be that when the radius of
QPO generation reaches some critical point, interactions with the stellar
magnetosphere or the properties of the boundary layer change suddenly and
the quality factor drops.  However, one would expect that if the mass
accretion rate were to change significantly, the details of the plasma
interactions would also change, leading to a significant variation in the
quality factor with mass accretion rate.  The current evidence is that
the QPO properties do not correlate well with countrate, but countrate is
not a reliable measure of mass accretion rate (e.g., M\'endez et al.
2001).  If it is demonstrated that the dependence of quality
factor on frequency is roughly independent of mass accretion rate, then
these scenarios would be less likely.

An attractive option would then be that the root cause of the
drop in quality factor has to do with the geometry of the spacetime,
because this would depend directly on the radius of QPO generation and
only secondarily on details of plasma interactions.  This would also
explain the consistency of the saturation frequency and rms vs.
frequency curves over many countrates.  A natural candidate
for a mechanism is one related to the flattening of the specific angular
momentum curve with radius near the innermost stable circular orbit
(ISCO).  The gas  orbiting around the neutron star is subjected to a
number of stresses that remove angular momentum, hence close to the ISCO
the inward radial speed will tend to increase.  This can help define a
narrow range of radii in which the QPO is generated, which will tend to
increase the quality factor as the radius decreases (see Miller  et al.
1998).  This could explain the gradual rise in $Q$ with  frequency at
low frequencies.  However, if the range of radii covered during the
lifetime of the QPO becomes too large, the range of frequencies sampled
during its lifetime is also large, hence the quality factor drops
sharply when the radius is too close to the ISCO.  Because the upper QPO
has a lower quality factor than the lower QPO, the crossover point
(where radial drift becomes dominant) would be at smaller radius, hence
in this scenario it makes sense that the higher $Q$ oscillation would
show a drop in $Q$ at a lower frequency, whereas the upper QPO would
not necessarily show a clear drop.

If this interpretation is correct then one may use the sharp drop in
coherence to estimate the frequency of the ISCO.  We assume that $Q=0$
corresponds to the ISCO.  The frequency at which this will occur is
somewhat uncertain, because of the lack of a fundamental theory to
predict the functional dependence of $Q$ on frequency.  We will use a
linear extrapolation, which appears to fit the data adequately. This
gives us a maximum frequency $\nu_{\rm low,max}$ for the lower QPO.
We then assume, as in most models, that the upper QPO is at
approximately the orbital frequency at some radius in the disk and the
lower QPO is less than this by approximately the spin frequency or
half the spin frequency, depending on the source (Wijnands et al.
2003; see Lamb \& Miller 2005 for a theoretical discussion). We
therefore estimate the orbital frequency at the ISCO by adding the
difference frequency between the upper and lower QPOs, so $\nu_{\rm
ISCO}\approx \nu_{\rm low,max}+\Delta\nu$.  Using this approach, we
estimate $\nu_{\rm ISCO}\approx 1200-1300$~Hz in these three sources .
The mass then follows from
\begin{equation}
M/M_\odot=(\nu_{\rm ISCO}/2200~{\rm Hz})(1+0.75j)
\end{equation}
(Miller et al. 1998), where $j\equiv cJ/GM^2$ is the
dimensionless angular momentum, with $j\approx 0.1$ a
typical value.  Therefore, $\nu_{\rm ISCO}\approx 1200-1300$~Hz
implies $M\sim 1.8-2.0\,M_\odot$.  This is higher than inferred
for double neutron stars but consistent with modern equations
of state for matter beyond nuclear density (Lattimer \& Prakash 2001).

The ISCO interpretation of the sharp drop in quality factor is
subject to falsification in at least two ways.  First, the
interpretation is obviously incorrect if in a given source a
QPO is observed above the putative ISCO frequency.  The
uncertainty in extrapolation to $Q=0$ means that there is some
leeway, but not more than a few tens of Hertz.  The
second possibility is that a source shows behavior similar to
that described here, but at a frequency low enough that the
implied mass is higher than plausible for a neutron star.
From Lattimer \& Prakash (2001), $M<2.8\,M_\odot$ is required
even for mean field theories, so if the inferred $\nu_{\rm ISCO}<800$~Hz
for any source then this interpretation is called into question.
However, if neither of these falsifications occur and additional
sources show a sharp drop in $Q$ that depends only on frequency,
the data may indeed be showing evidence for the innermost stable
circular orbit.

\section*{Acknowledgments}
MCM was supported in part by a National Research Council fellowship
at Goddard Space Flight Center. This research has made use of data
obtained from the High Energy Astrophysics Science Archive Research
Center (HEASARC), provided by NASA's Goddard Space Flight Center.


\begin{thebibliography}{}

\bibitem[\protect\citeauthoryear{}{}]{akmal98} Akmal A., Pandharipande V.
R., Ravenhall D. G., 1998, PhRvC, 58, 1804

\bibitem[\protect\citeauthoryear{Barret et
al.}{2005}]{2005MNRAS.357.1288B}  Barret D., Klu{\' z}niak W., Olive J.~F.,
Paltani S., Skinner G.~K., 2005,  MNRAS, 357, 1288

\bibitem[\protect\citeauthoryear{Barret, Olive, \&
Miller}{2005}]{2005MNRAS.361..855B} Barret D., Olive J.-F., Miller M.~C.,
2005, MNRAS, 361, 855

\bibitem[\protect\citeauthoryear{Berger et al.}{1996}]{berger96} Berger,
M.~et al.\ 1996, \apjl, 469, L13

\bibitem[\protect\citeauthoryear{Ford et al.}{1998}]{1998ApJ...508L.155F}
Ford E.~C., van der Klis M., van Paradijs J., M{\' e}ndez M., Wijnands R.,
Kaaret P., 1998, ApJ, 508, L155

\bibitem[\protect\citeauthoryear{Kluzniak, Michelson, \&
Wagoner}{1990}]{1990ApJ...358..538K} Klu\'zniak W., Michelson P., Wagoner
R.~V., 1990, ApJ, 358, 538

\bibitem[\protect\citeauthoryear{Kluzniak \&
Wagoner}{1985}]{1985ApJ...297..548K} Klu\'zniak W., Wagoner R.~V., 1985,
ApJ,  297, 548

\bibitem[\protect\citeauthoryear{Lamb \& Miller}{2005}]{2005ApJ Submitted}
Lamb, F.~K., Miller, M.~C. 2005, ApJ, submitted (astro-ph/0308179)

\bibitem[\protect\citeauthoryear{Lattimer \&  Prakash}{2001}]{lattimer01}
Lattimer J.~M., Prakash M., 2001, ApJ,  550, 426

\bibitem[\protect\citeauthoryear{M{\' e}ndez, van der Klis, \& van
Paradijs}{1998a}]{1998ApJ...506L.117M} M{\' e}ndez M., van der Klis M., van
Paradijs J., 1998a, ApJ, 506, L117

\bibitem[\protect\citeauthoryear{M{\' e}ndez et al.}{1998b}]{mendez98b}
M{\' e}ndez, M., et al.\ 1998b, \apjl, 494, L65

\bibitem[\protect\citeauthoryear{M{\' e}ndez et al.}{1999}]{mendez99} M{\'
e}ndez M., van der Klis M., Ford  E.~C., Wijnands R., van Paradijs J.,
1999, ApJ, 511, L49

\bibitem[\protect\citeauthoryear{M{\' e}ndez, van der Klis, \&
Ford}{2001}]{mendez01} M{\' e}ndez M., van der Klis M., Ford  E.~C., 2001,
ApJ, 561, 1016

\bibitem[\protect\citeauthoryear{Miller, Lamb, \& Psaltis}{1998}]{miller98}
Miller, M.~C., Lamb, F.~K., \& Psaltis, D.\ 1998, \apj, 508, 791

\bibitem[\protect\citeauthoryear{Wijnands et al.}{2003}]
{2003Nature...424..44} Wijnands R., van der Klis M., Homan, J.,
Chakrabarty, D., Markwardt, C.~B., Morgan, E.~H. 2003, Nature, 424, 44

\bibitem[\protect\citeauthoryear{Wijnands et
al.}{1998}]{1998ApJ...495L..39W} Wijnands R., van der Klis M., Mendez M.,
van Paradijs J., Lewin W.~H.~G., Lamb F.~K., Vaughan B., Kuulkers E.,
1998,  ApJ, 495, L39

\end{thebibliography}
\end{document}